\documentclass[prd,twocolumn,superscriptaddress,nofootinbib]{revtex4-2}

\usepackage{graphicx}
\usepackage{dcolumn}
\usepackage{bm}
\usepackage{amssymb}
\usepackage{amsfonts}
\usepackage{latexsym}
\usepackage[T1]{fontenc}
\usepackage{epsf,amsfonts}
\usepackage{epsfig}
\usepackage{amsthm}
\usepackage{amsmath}
\usepackage{color}
\usepackage{slashed}
\usepackage{mathrsfs}
\usepackage{float}
\usepackage{esvect}
\usepackage{mathtools}
\usepackage{bbm}
\usepackage[usenames,dvipsnames]{xcolor}
\usepackage[colorlinks=true,citecolor=Blue,linkcolor=RubineRed,urlcolor=Blue]{hyperref}
\usepackage{subcaption}  

\newcommand{\sumint}{\mathop{\mathrlap{\sum}\int}}

\begin{document}

\title{Positivity bounds from thermal field theory entropy}

\author{Xin-Yi Liu}
\email{liuxinyi23@mails.ucas.ac.cn}
\affiliation{School of Fundamental Physics and Mathematical Sciences, Hangzhou Institute for Advanced Study, UCAS, Hangzhou 310024, China}
\affiliation{Institute of Theoretical Physics, Chinese Academy of Sciences, Beijing 100190, China}
\affiliation{University of Chinese Academy of Sciences (UCAS), Beijing 100049, China}
\author{Yongjun Xu}
\email{xuyongjun23@mails.ucas.ac.cn}
\affiliation{School of Fundamental Physics and Mathematical Sciences, Hangzhou Institute for Advanced Study, UCAS, Hangzhou 310024, China}
\affiliation{Institute of Theoretical Physics, Chinese Academy of Sciences, Beijing 100190, China}
\affiliation{University of Chinese Academy of Sciences (UCAS), Beijing 100049, China}

\begin{abstract}
We present an approach to deriving positivity bounds on effective field theories by analyzing the thermodynamic behavior of thermal quantum field systems. Focusing on scalar theories with higher-dimensional operators, we compute the finite-temperature entropy using thermal field theory techniques. We argue that consistency with fundamental thermodynamic principles---specifically, the expectation that entropy increases with the introduction of new degrees of freedom---imposes nontrivial constraints on Wilson coefficients. In particular, we show that the coefficient of the leading dimension-8 operator must be strictly positive. This thermodynamic perspective offers an alternative to traditional S-matrix-based derivations of positivity bounds and provides a complementary perspective on the interplay between entropy, unitarity, and causality in quantum field theory.
\end{abstract}

\maketitle

\section{Introduction}

Effective Field Theory (EFT) provides a systematic framework for describing
low-energy phenomena in systems with a hierarchy of scales. It is especially
useful when the ultraviolet (UV) completion is unknown or strongly coupled, as
often occurs in gravity and hadronic physics. In practice, EFTs are built in two
complementary ways: a top-down approach, which integrates out heavy degrees of
freedom from a known UV theory, and a bottom-up approach, which enumerates all
operators consistent with low-energy symmetries and treats their coefficients as
free parameters.

However, if we assume the existence of a consistent UV completion that respects causality, locality, and unitarity, then the low-energy Wilson coefficients are not entirely arbitrary \cite{Adams:2006sv}. Constraints arising from these fundamental principles—commonly referred to as positivity bounds—have traditionally been derived using analyticity properties of scattering amplitudes and dispersion relations \cite{Bellazzini:2019xts,Arkani-Hamed:2021ajd,Henriksson:2021ymi,Alberte:2020bdz,Saraswat:2016eaz,Guerrieri:2021ivu,Pham:1985cr, Pennington:1994kc,Nicolis:2009qm, Komargodski:2011vj,Remmen:2019cyz,Herrero-Valea:2019hde,Bellazzini:2017fep,deRham:2017avq,deRham:2017zjm,deRham:2017imi,Wang:2020jxr, Tokuda:2020mlf,Li:2021lpe,Caron-Huot:2021rmr,Du:2021byy,Bern:2021ppb,Li:2022rag, Caron-Huot:2022ugt,Herrero-Valea:2020wxz,EliasMiro:2022xaa,Bellazzini:2021oaj, Sinha:2020win, Trott:2020ebl,Herrero-Valea:2022lfd,Hong:2023zgm,Chiang:2022jep,Huang:2020nqy,Noumi:2021uuv, Xu:2023lpq, Chen:2023bhu,Noumi:2022wwf,deRham:2022hpx, Hong:2024fbl,Bern:2022yes,Ma:2023vgc,DeAngelis:2023bmd,Acanfora:2023axz,Aoki:2023khq,Xu:2024iao,EliasMiro:2023fqi,McPeak:2023wmq,Riembau:2022yse,Caron-Huot:2024tsk,Caron-Huot:2024lbf,Wan:2024eto,Buoninfante:2024ibt,Berman:2024owc,deRham:2025vaq,Berman:2025owb}.

In this work we pursue an alternative route. Motivated by studies relating black
hole thermodynamics to quantum-gravity consistency conditions
\cite{Arkani-Hamed:2006emk,Cheung:2018cwt,Cottrell:2016bty,Hebecker:2017uix,Abe:2023anf,DeLuca:2022tkm,Barbosa:2025uau,Cao:2022ajt,Cao:2022iqh,Aalsma:2019ryi,Hamada:2018dde},
we ask whether analogous constraints can be obtained directly in finite-
temperature quantum field theory. Our central proposal is that thermodynamic
consistency---specifically, requiring that the thermal entropy does not decrease
when additional microscopic degrees of freedom are integrated out---imposes
positivity constraints on EFT Wilson coefficients.

To make this idea concrete, we analyze a shift-symmetric scalar EFT, compute its
finite-temperature entropy, and show that the leading dimension-8 coefficient
must be strictly positive. This identifies a direct link between thermal
statistical mechanics and EFT consistency, providing a complement to standard
amplitude-based positivity arguments.

\section{Basic idea}

Consider a system in thermal equilibrium characterized by two widely separated
mass scales: a heavy mass $M$ and a light mass $m$. We focus on an intermediate
temperature regime
\begin{equation}
    m \ll T \ll M ,
\end{equation}
in which thermal excitations of the heavy degrees of freedom are exponentially
suppressed, while the light modes remain thermally active. In this regime, the thermal entropy of the full theory can be well approximated by that of an EFT containing only the light degrees of freedom. The effects of heavy degrees of freedom are encoded in higher-dimensional interaction terms that arise from integrating out the heavy fields.
\begin{equation}\label{eq1}
    S_{m+M,\mathrm{thermal}} \simeq S_{m\text{-EFT},\mathrm{thermal}} .
\end{equation}

A naive classical intuition suggests that introducing additional microscopic
degrees of freedom should increase the number of accessible states and hence the
thermal entropy. This motivates the inequality
\begin{equation}\label{eq2}
    S_{m+M,\mathrm{thermal}} > S_{m\text{-alone},\mathrm{thermal}} ,
\end{equation}
where $S_{m\text{-alone},\mathrm{thermal}}$ denotes the thermal entropy of a free
theory containing only the light sector.

However, this intuition does not directly carry over to quantum theory when
entropy is defined as the von Neumann entropy of a reduced density matrix, due to
the presence of quantum entanglement. For example, a bipartite system $A+B$
prepared in a pure state has vanishing total entropy,
\begin{equation}
    S_{A+B}=0 ,
\end{equation}
while the reduced state of subsystem $A$ obtained by tracing out $B$ is generally
mixed and has nonzero entropy,
\begin{equation}\label{eq3}
    S_A > 0 .
\end{equation}

In the present context, tracing out heavy degrees of freedom therefore introduces
entanglement contributions that must be treated with care. To make this explicit,
let $S_m$ and $S_M$ denote the von Neumann entropies of the reduced density
matrices of the light and heavy sectors, respectively, obtained from the full
theory. Schematically, one may write
\begin{align}
    S_m &\simeq S_{m,\mathrm{thermal}} + S_{m\rightarrow M,\mathrm{ent}} ,
    \label{eq4}\\
    S_M &\simeq S_{M,\mathrm{thermal}} + S_{M\rightarrow m,\mathrm{ent}} ,
    \label{eq5}
\end{align}
where $S_{m,\mathrm{thermal}}$ and $S_{M,\mathrm{thermal}}$ denote the thermal free 
(entropy-from-occupation) contributions, while
$S_{m\rightarrow M,\mathrm{ent}}$ and $S_{M\rightarrow m,\mathrm{ent}}$ encode
entanglement between the two sectors induced by interactions between the two
fields. We stress that Eqs.~\eqref{eq4} and
\eqref{eq5} are not exact additive identities for von Neumann entropy; rather,
they represent an approximate separation that is valid in the weakly coupled
regime $m \ll T \ll M$, where thermal and entanglement contributions can be
parametrically distinguished.

In quantum field theory, entanglement entropy is generally ultraviolet divergent and requires a regulator, reflecting short-distance correlations across the entangling surface. A standard regularization, familiar for example from conformal field theory, introduces a short-distance cutoff $\epsilon$, which gives rise to local area-law divergences that are independent of temperature. The entanglement entropy in our setup behaves in the same way. However, since our analysis only involves differences of entropies, these divergent contributions cancel in the final result.

More precisely, consider the low-temperature regime $T \ll M$. In this limit, the thermal entropy associated with the heavy sector is exponentially suppressed,
\begin{equation}
    S_{M,\mathrm{thermal}} \propto T^{3/2} e^{-M/T},
\end{equation}
and can therefore be neglected. Consequently, the entropy $S_M$ is dominated by entanglement contributions and can be approximated as
\begin{equation}
    S_M \simeq S_{M \rightarrow m,\mathrm{ent}} .
\end{equation}
Taking the difference between $S_m$ and $S_M$ isolates the thermal contribution of the light degrees of freedom,
\begin{equation}\label{eq7}
    S_{m,\mathrm{thermal}} \simeq \bigl| S_m - S_M \bigr|_{\,T \ll M}.
\end{equation}

We now invoke the Araki--Lieb inequality~\cite{Araki:1970ba,Witten:2018zva}, which
holds for the von Neumann entropies of any bipartite quantum system,
\begin{equation}\label{eq8}
    S_{m+M} \ge \bigl| S_m - S_M \bigr| .
\end{equation}
Applying this inequality in the regime $m \ll T \ll M$, and combining it with
Eq.~\eqref{eq1}, we obtain
\begin{equation}
    S_{m,\mathrm{EFT}} > S_{m,\mathrm{thermal}} .
\end{equation}
This result sharpens the heuristic expectation expressed in
Eq.~\eqref{eq2} by accounting for entanglement effects and regulator-dependent
contributions in a controlled manner. It provides the thermodynamic inequality
that underlies our subsequent entropy-based positivity bound.\footnote{Although one might initially expect $S_{m,\text{EFT}}$ and $S_{m,\text{thermal}}$ to coincide, there is a crucial distinction between tracing out heavy degrees of freedom and integrating them out in the effective field theory (EFT). In constructing $S_m$, we perform a partial trace over the heavy sector $M$, retaining only the light degrees of freedom with mass $m$. In contrast, when evaluating $S_{m,\text{EFT}}$, we integrate out not only the heavy fields with mass $M$ but also the high-energy modes of the light field itself. Consequently, $S_m$ and $S_{m,\text{EFT}}$ are not identical: the latter includes additional contributions from integrating over the UV modes of the light field.}
\section{Calculation of Thermal entropy}
We begin by considering a simple scalar field theory that possesses a shift symmetry, which physically corresponds to taking the massless limit $(m \to 0)$. Such a field often emerges as a Goldstone boson in effective field theories, representing the low-energy fluctuations around spontaneously broken continuous symmetries.  Throughout this section, we adopt natural units with \(k_B=\hbar=1\).

The low-energy effective Lagrangian for such a scalar field \(\phi\), truncated at dimension eight, is
\begin{align}
    \mathcal{L}_M=-\frac12\left(\partial\phi\right)^2+\frac{c}{M^4}\left(\partial\phi\right)^4+\cdots.
\end{align}
where \(c\) is a dimensionless coupling constant, and \(M\) denotes a high-energy cutoff scale in the effective theory. The first term corresponds to the standard kinetic energy of a free scalar field, while the second term represents the leading-order self-interactions of the field.

To incorporate finite-temperature effects, we consider the system in thermal 
equilibrium with a heat bath at temperature \(T\), which allows us to compute the 
statistical properties of the scalar field within a consistent quantum field 
theoretical framework.

From a theoretical perspective, to prepare a thermal equilibrium state at 
temperature \(T = \tfrac{1}{\beta}\), we can employ the Euclidean path integral 
formalism on a compact manifold with temporal circumference \(\beta\). 
Concretely, this approach involves the following steps \cite{Laine:2016hma}:
\begin{itemize}
    \item Imaginary Time (Wick Rotation): We replace real time \(t\) by imaginary (Euclidean) time \(\tau\) through the transformation:
    \begin{align}
        \tau\equiv it.
    \end{align}
    This step transforms the Minkowski spacetime into a Euclidean spacetime, making the field theory resemble a classical statistical system.
    \item Euclidean Lagrangian: Under Wick rotation, the Lagrangian transforms into its Euclidean counterpart. The Euclidean Lagrangian \(L_E\) is related to the original Minkowski Lagrangian by:
    \begin{align}
        L_E(\tau,x)\equiv-\mathcal{L}_M(t,x),
    \end{align}
    \item Periodic Boundary Conditions and Temperature: At finite temperature, the Euclidean time direction becomes compactified with period \(\beta\), where
    \begin{align}
        \tau \sim \tau+\beta, \text{ where }\beta\equiv\frac{1}{T}.
    \end{align}
    This periodicity reflects thermal equilibrium and emerges naturally from the trace over states in the partition function:
    \begin{align}
        \mathcal{Z}(T)=\text{Tr}(e^{-\beta H}).
    \end{align}
\end{itemize}
Following these steps, the Euclidean action for the scalar theory becomes:
\begin{align} S_E&=S_0+S_I\notag\\
&=\int^\beta_0d\tau\int d^3\mathbf{x}\,\left[\frac12\left(\partial\phi\right)^2-\frac{c}{M^4}\left(\partial\phi\right)^4+\cdots\right]\notag\\
&=\int_X\left[\frac12\left(\partial\phi\right)^2-\frac{c}{M^4}\left(\partial\phi\right)^4+\cdots\right]
\end{align}
where we clearly separate the free action \(S_0\) (quadratic in \(\phi\)) from the interaction terms \(S_I\) (higher-order terms in \(\phi\)). Note that in the zero-temperature limit \((\beta\to\infty)\), the periodicity condition disappears, and we naturally recover the standard Euclidean quantum field theory defined on infinite spacetime.

A crucial motivation for studying this theory at finite temperature is understanding how thermal entropy, a fundamental quantity describing the disorder and microscopic states of the system, constrains the properties of the theory. In particular, as we shall see from detailed calculations, consistency conditions related to the entropy demand the positivity of certain coupling constants. In this model, we will see that this implies:
\begin{align}
    c>0.
\end{align}
Thus, we can find that thermal field theory not only provides a rich physical framework to study quantum systems at finite temperature but also imposes powerful theoretical constraints on allowed interactions.

To quantify these ideas precisely, we construct the thermal partition function:
\begin{align}
    \mathcal{Z}(T) =C \int \mathcal{D} \phi e^{-S_E}=C \int \mathcal{D} \phi e^{\left(-S_0-S_{I}\right)},
\end{align}
where the constant \(C\) contributes only to the vacuum energy. Given that the coupling \(\frac{c}{M^4}\) is small (since \(M\) is large compared to other physical scales in the theory), we can expand the partition function perturbatively around the free theory. Explicitly, we obtain:
\begin{align}\label{PartitionFunc}
    \mathcal{Z}(T) & =C \int \mathcal{D} \phi e^{-S_0}\left[1-S_{\mathrm{I}}+\frac{1}{2} S_{\mathrm{I}}^2-\frac{1}{6} S_{\mathrm{I}}^3+\cdots\right] \notag\\ & =\mathcal{Z}_{(0)}\left[1-\left\langle S_{\mathrm{I}}\right\rangle_0+\frac{1}{2}\left\langle S_{\mathrm{I}}^2\right\rangle_0-\frac{1}{6}\left\langle S_{\mathrm{I}}^3\right\rangle_0+\cdots\right] ,
\end{align}
where
\begin{align}
&\mathcal{Z}_{(0)} \equiv C \int \mathcal{D} \phi e^{-S_0},\\
&\langle\cdots\rangle_0 \equiv \frac{\int \mathcal{D} \phi[\cdots] e^{-S_0}}{\int \mathcal{D} \phi e^{-S_0}}.
\end{align}

The free energy density \(f(T)\) is defined in terms of the thermodynamic free energy as
\begin{align}
    f(T)\equiv\lim_{V\to\infty}\frac{F(T,V)}{V},
\end{align}
with \(F(T,V)=-T\ln\mathcal{Z}\) and
\begin{align}
    \frac{F(T,V)}{V}=\frac{F_{(0)}}{V}-\frac{T}{V}\left[-\langle S_I\rangle_0+\frac{1}{2}\left(\langle S_I^2\rangle_0-\langle S_I\rangle^2_0\right)-\cdots\right].
\end{align}
It is standard to separate the free energy into contributions from the free theory and corrections due to interactions:
\begin{align}
    f=f_0+f_{\left(\ge1\right)}.
\end{align}
For leading-order and first-order, we have:
\begin{align}
    f_{(0)}&=\lim_{V\to\infty}\frac TVS_0=\lim_{V\to\infty}\frac{-T\ln\mathcal{Z}_0}{V},\\
    f_{(1)}&=\lim_{V\to\infty}\frac TV\langle S_I\rangle_0\notag\\
    &=\lim_{V\to\infty}\!\!-\frac TV\!{\int_X}\!\frac{c}{M^4}\!\langle\partial_\mu\phi(X)\partial^\mu\phi(X)\partial_\nu\phi(X)\partial^\nu\phi(X)\rangle_0\label{f1}.
\end{align}
\begin{figure}[htbp]
    \centering
    \begin{subfigure}[t]{0.36\linewidth}
        \centering
        \includegraphics[width=\linewidth]{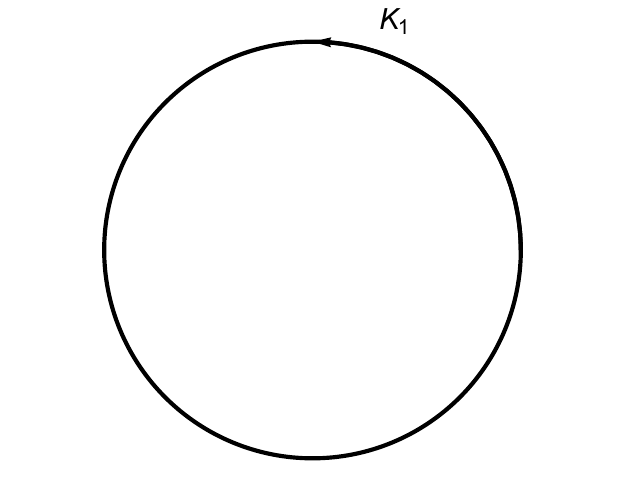}
        \caption{\(f_{(0)}\)}
        \label{f0&f1:f0}
    \end{subfigure}
    \hspace{0.01\linewidth}
    \begin{subfigure}[t]{0.6\linewidth}
        \centering
        \includegraphics[width=\linewidth]{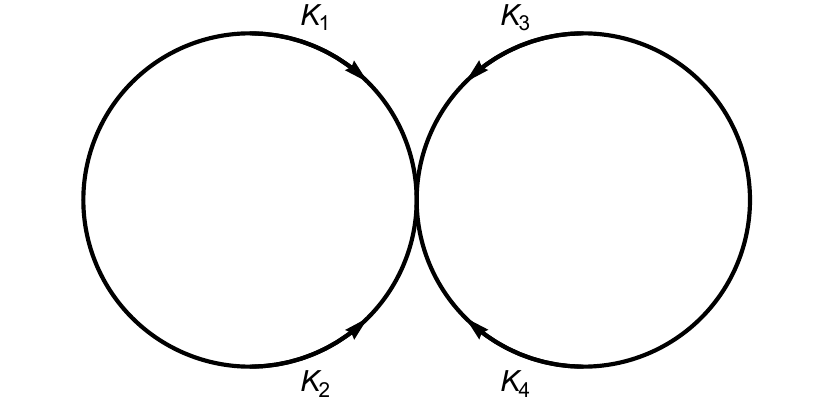}
        \caption{\(f_{(1)}\)}
        \label{f0&f1:f1}
    \end{subfigure}
    \caption{Diagrams contributing to the free-energy expansion. Panel (b) shows one representative contraction, \(K_1\) with \(K_2\); the \(K_1\)-\(K_3\) and \(K_1\)-\(K_4\) contractions contribute analogously.
    }
    \label{f0&f1}
\end{figure}
These two quantities correspond to Fig.~\ref{f0&f1}. Before carrying out the explicit calculation, it is useful to estimate their scaling. The leading term \(f_{(0)}\) involves a single momentum integral and scales as \(T^4\). For \(f_{(1)}\), Fig.~\ref{f0&f1:f1} contains two independent loop momenta, and dimensional analysis gives \(f_{(1)}\sim \frac{c}{M^4}T^8\). 

We now proceed to compute \(\mathcal{Z}_0\) from which \(f_{(0)}\) can be extracted. The partition function \(\mathcal{Z}_0\) takes the form:
\begin{align}
\mathcal{Z}_0 &= C\int \mathcal{D}\phi \exp\left[-\int_X \frac{1}{2}(\partial \phi)^2\right].
\end{align}
We adopt the following Fourier expansion for the scalar field:
\begin{align}
    \phi\left(X\right)= \sqrt{\beta V}\sum_{\omega_n}\int\frac{d^d\textbf{k}}{\left(2\pi\right)^d}\tilde{\phi}(K)e^{iK\cdot X}\equiv\sumint _K \tilde{\phi}(K) e^{i K \cdot X},
\end{align}
where \(K\equiv(\omega_n,\textbf{k})\), with bosonic Matsubara frequencies \(\omega_n=2\pi T n,\ n\in\mathbb{Z}\), imposed by the periodicity \(e^{i\omega_n\beta}=1\) \cite{Matsubara:1955ws}. Furthermore, for the real scalar field, we have the condition: \(\phi^\ast(X)=\phi(X)\Rightarrow\tilde\phi^\ast(K)=\tilde\phi(-K)\). Substituting this into the action gives:
\begin{align}
    \frac{\mathcal{Z}_0}{C} &= \int \mathcal{D}\phi \exp\left[\int_X \frac12\sumint_{K,K^\prime}(K_\mu K^{\prime\mu})\right.\notag\\
    &\left.\tilde{\phi}(K)\tilde{\phi}(K')\times e^{i K\cdot X + i K^\prime\cdot X}\right]\notag\\
    &=\int \mathcal{D}\phi \exp\left[-\frac12\beta V\sum_{\omega_n}\int\frac{d^3\mathbf{k}}{(2\pi)^3}K^2\tilde\phi(-K)\tilde\phi(K)\beta\right]\notag\\
    &=\int \mathcal{D}\phi \exp\left[-\frac12\beta^2V\sum_{\omega_n}\int\frac{d^3\mathbf{k}}{(2\pi)^3}K^2\tilde\phi^\ast(K)\tilde\phi(K)\right]\notag\\
    &=\int\mathcal{D}\phi\exp\left[\sum_{K}\left(-\frac12\tilde\phi^\ast(K) K^2\beta^2\tilde\phi(K)\right)\right].
\end{align}
Here we used \(\int^\beta_0d\tau\, e^{i(\omega_n+\omega_n^{\prime})\tau}=\beta\,\delta(\omega_n+\omega_n^{\prime})\) and \(\int d^3\mathbf{x}\,e^{i(\mathbf{k}+\mathbf{k^\prime})\mathbf{x}}=(2\pi)^3\delta^{(3)}(\mathbf{k}+\mathbf{k^\prime})\). In the last equality, \(\sum_{K}\) denotes \(V\sum_{\omega_n}\int\frac{d^3\mathbf{k}}{(2\pi)^3}\).
Using the Gaussian integral identity:$\int d^D x e^{-\frac{1}{2} \mathbf{x}^\mathbf{T}\hat{A} \mathbf{x}}=(2 \pi)^{D / 2}(\operatorname{det} \hat{A})^{-1 / 2}$,
we obtain: 
\begin{align}
    \mathcal{Z}_0=C\cdot\sqrt{\frac{1}{\det(K^2\beta^2)}},
\end{align}
Since the constant \(C\) is independent of temperature, we drop it in what follows. Therefore,
\begin{align}
    \ln\mathcal{Z}_0&=-\frac{1}{2}\ln\det(K^2\beta^2),\notag\\
    &=-\frac{1}{2}\sum_K\ln \left(K^2\beta^2\right)=-\frac{1}{2}V\int\frac{d^3 \mathbf{k}}{(2\pi)^3}\sum_{\omega_n}\ln (K^2\beta^2),\notag\\   
    &=V\int\frac{d^3 \mathbf{k}}{(2\pi)^3}\left[-\frac{1}{2}\beta|\mathbf{k}|-\ln\left(1-e^{-\beta|\mathbf{k}|}\right)\right].
\end{align}
where the last step uses the standard Matsubara sum formula \eqref{MS2}.
Thus, the free energy density becomes:
\begin{align}
    \int\frac{d^3\mathbf{k}}{(2\pi)^3}\left[\frac{1}{2}|\mathbf{k}|+T\cdot\ln\left(1-e^{-|\mathbf{k}|/T}\right)\right].
\end{align}
The leading temperature-independent term reflects the zero-temperature vacuum energy, which lacks thermal significance. As such, it can be consistently subtracted when evaluating thermodynamic quantities. This yields the renormalized result:
\begin{align}
    f_{(0)}&=\int\frac{d^3\mathbf{k}}{(2\pi)^3}T\cdot\ln\left(1-e^{-|\mathbf{k}|/T}\right),\notag\\
    &=\frac{T}{2\pi^2}\int dk \cdot k^2\ln\left(1-e^{-k/T}\right)\notag\\
    &=\frac{T^4}{2\pi^2}\int dx \cdot x^2\ln\left(1-e^{-x}\right),\notag\\
    &=-\frac{\pi^2T^4}{90}.
\end{align}
In the third line we performed the change of variable \(k/T\to x\). We now turn to
the evaluation of \(f_{(1)}\). From the last line of Eq.~\eqref{f1},
\(\langle\partial_\mu\phi(X)\partial^\mu\phi(X)\partial_\nu\phi(X)\partial^\nu\phi(X)\rangle_0\)
is independent of spacetime coordinates by translational invariance, so the
spacetime integral is trivial. Therefore,
\begin{align}
    f_{(1)}=-\frac{c}{M^4}\langle\partial_\mu\phi(0)\partial^\mu\phi(0)\partial_\nu\phi(0)\partial^\nu\phi(0)\rangle_0.
\end{align}
where the prefactor \(\frac{T}{V}\) in Eq.~\eqref{f1} is canceled by
\(\frac{T}{V}\int_X=\frac{T}{V}\beta V=1\). To handle the derivatives, we perform a Fourier transform of the fields:
\begin{align}
    f_{(1)}&=-\frac{8c}{M^4}\langle\sumint_{K_1}\ iK_{1,\mu}\tilde{\phi}\left(K_1\right) \sumint_{K_2}\ iK_{2}^\mu\tilde{\phi}\left(K_2\right)\notag\\
    &\times \sumint_{K_3}\ iK_{3,\nu}\tilde{\phi}\left(K_3\right) \sumint_{K_4}\ iK_{4}^\nu\tilde{\phi}\left(K_4\right)\\
    &+\left(2\leftrightarrow3\right)+\left(2\leftrightarrow4\right)\rangle_0\notag\\
    &=-\frac{8c}{M^4}\!\!\!\!\!\!\!\!\sumint_{K_1,K_2,K_3,K_4}\!\!\!\!\!\!\!\!\left(K_{1,\mu}K_2^\mu K_{3,\nu}K_4^\nu+\left(2\leftrightarrow3\right)+\left(2\leftrightarrow4\right)\right)\notag\\
    &\times \langle\tilde{\phi}(K_1)\tilde{\phi}(K_2)\tilde{\phi}(K_3)\tilde{\phi}(K_4)\rangle_0.
\end{align}
Using Wick’s theorem, we have:
\begin{align}
    f_{(1)}=&-\frac{c}{M^4}\sumint_{K_1,K_2,K_3,K_4}K_{1,\mu}K_2^\mu K_{3,\nu}K_4^\nu\notag\\
    &\times{8}\left[\langle\tilde\phi(K_1)\tilde\phi(K_2)\rangle_0\langle\tilde\phi(K_3)\tilde\phi(K_4)\rangle_0\right.\notag\\
    &\left.+\langle\tilde\phi(K_1)\tilde\phi(K_3)\rangle_0\langle\tilde\phi(K_2)\tilde\phi(K_4)\rangle_0\right.\notag\\
    &\left.+\langle\tilde\phi(K_1)\tilde\phi(K_4)\rangle_0\langle\tilde\phi(K_2)\tilde\phi(K_3)\rangle_0\right].
    \end{align}
With \(\langle\tilde\phi(K_1)\tilde\phi(K_2)\rangle_0=\frac{T^2}{V}\delta_{\omega_{1_n}+\omega_{2_n},0}(2\pi)^d\delta^{(d)}(\mathbf{k_1}+\mathbf{k_2})\frac{1}{K_1^2}\), we obtain:
\begin{align}\label{f1-2}
    f_{(1)}=&-\frac{{8\,}c}{M^4}T^2\sum_{\omega_{1_n}\omega_{3_n}}\int\frac{d^3\mathbf{k_1}}{(2\pi)^3}\frac{d^3\mathbf{k_3}}{(2\pi)^3}K_1^2K_3^2\frac{1}{K_1^2}\frac{1}{K_3^2}\notag\\
    &-\frac{{16\,}c}{M^4}T^2\sum_{\omega_{1_n}\omega_{2_n}}\int\frac{d^3\mathbf{k_1}}{(2\pi)^3}\frac{d^3\mathbf{k_2}}{(2\pi)^3}(K_1\cdot K_2)^2\frac{1}{K_1^2}\frac{1}{K_2^2}\notag\\
    =&-\frac{{8\,}c}{M^4}T^2\sum_{\omega_{n}\omega_{n}^\prime}\int\frac{d^3\mathbf{k}}{(2\pi)^3}\frac{d^3\mathbf{k^\prime}}{(2\pi)^3}\cdot\mathbf{1}\notag\\
    &-\frac{{16\,}c}{M^4}T^2\!\!\sum_{\omega_{n}\omega_{n}^\prime}\!\!\int\!\!\frac{d^3\mathbf{k}}{(2\pi)^3}\!\frac{d^3\mathbf{k^\prime}}{(2\pi)^3}\!(K\!\cdot\! K^\prime)^2{\frac{1}{\omega_n^2+\mathbf{k}^2}\frac{1}{\omega_n^{\prime 2}+\mathbf{k}^{\prime2}}}.
\end{align}
where we relabeled the four external momenta into two independent ones,
\(K=(\omega_n,\mathbf{k})\) and \(K^\prime=(\omega_n^\prime,\mathbf{k^\prime})\), as shown in Fig.~\ref{f0&f1:f1}.


The first term in Eq.~\eqref{f1-2} is proportional to the constant sum-integral
\begin{align}\label{integralI}
\mathcal{I}\equiv 
T^2 \sum_{\omega_n,\omega_n'} 
\int \frac{d^3 \mathbf{k}}{(2\pi)^3}
\int \frac{d^3 \mathbf{k}'}{(2\pi)^3} 1
=
\left[
T \sum_{\omega_n} 
\int \frac{d^3 \mathbf{k}}{(2\pi)^3} 1
\right]^2 .
\end{align}
Although this expression carries an explicit factor of $T^2$, it does not correspond to a genuine thermal contribution. In thermal field theory, temperature dependence arises from the difference between the discrete Matsubara sum and the corresponding zero-temperature frequency integral. For a sufficiently regular function $F(\omega)$ one may write
\begin{align}
T\sum_n F(\omega_n)
=
\int \frac{d\omega}{2\pi}F(\omega)
+
\Delta_T[F],
\end{align}
where the first term, which is independent of $T$, represents the vacuum contribution and $\Delta_T[F]$ contains the genuine thermal correction.

For the present case the integrand is constant, $F(\omega)=1$, so the thermal
part vanishes. A convenient derivation follows from the Poisson summation
formula. To this end, we introduce a regulator
\(F_{\alpha}(\omega)=e^{-\alpha|\omega|}\) with \(\alpha>0\), which approaches
unity in the limit \(\alpha\to 0^+\). Applying the Poisson summation formula
\eqref{PoissonSumFormula} gives
\begin{align}
    T \sum_{n  \in \mathbb{Z}} F_\alpha\left(\omega_n\right)=\sum_{m \in \mathbb{Z}} \int \frac{d \omega}{2 \pi} e^{i m \beta \omega} F_\alpha(\omega) .
\end{align}
The \(m=0\) term reproduces the vacuum integral \(\int \frac{d\omega}{2\pi}F_\alpha(\omega)\). The remaining \(m\neq 0\) terms are proportional to \(\alpha/\left(\alpha^2+m^2\beta^2\right)\) and vanish as \(\alpha\to 0^+\). Hence \(\Delta_T[1]=0.\) (see Appendix~\ref{Poisson} for more details)

Therefore, the sum-integral in Eq.~\eqref{integralI} contains only the vacuum
contribution. Since physical thermodynamic quantities are defined after
subtracting the zero-temperature vacuum contribution (equivalently, by absorbing
it into a vacuum-energy counterterm), this term does not contribute to the
renormalized free energy or entropy and can be discarded. We thus retain only
the second term in Eq.~\eqref{f1-2}.
\begin{align}
    f_{(1)}\!=&-\!\frac{{16\,}c}{M^4}T^2\!\!\sum_{\omega_{n}\omega_{n}^\prime}\!\int\!\!\frac{d^3\mathbf{k}}{(2\pi)^3}\!\frac{d^3\mathbf{k^\prime}}{(2\pi)^3}(K\!\cdot\! K^\prime)^2\!\frac{1}{\omega_n^2+\mathbf{k}^2}\!\frac{1}{\omega_n^{\prime 2}+\mathbf{k}^{\prime2}}\notag\\
    =&-\!\frac{{16\,}c}{M^4}T^2\!\!\int\!\frac{d^3\mathbf{k} d^3\mathbf{k^\prime}}{(2\pi)^6}\!\!\sum_{\omega_n,\omega_n^\prime}\!\frac{\omega_n^2\omega_n^{\prime 2}{+2\omega_n\omega^\prime_n\mathbf{k}\!\cdot\!\mathbf{k^\prime}\!+\!\left(\mathbf{k}\!\cdot\!\mathbf{k^\prime}\right)^2}}{(\omega_n^2+\mathbf{k}^2)(\omega_n^{\prime 2}+\mathbf{k}^{\prime2})}\notag\\
    =&-\!\frac{{16\,}c}{M^4}T^2\!\!\int\!\frac{d^3\mathbf{k} d^3\mathbf{k^\prime}}{(2\pi)^6}\!\!\sum_{\omega_n,\omega_n^\prime}\!\frac{\omega_n^2\omega_n^{\prime 2}\!+\!{\left(\mathbf{k}\!\cdot\!\mathbf{k^\prime}\right)^2}}{(\omega_n^2+\mathbf{k}^2)(\omega_n^{\prime 2}+\mathbf{k}^{\prime2})}.
\end{align}
The middle cross term in the second line vanishes because it is odd under the integral. The remaining two numerator terms are treated separately: we first carry out the Matsubara sums and then perform the momentum integrals.
For the first term in the numerator,
\begin{align}\label{m1}
    &\sum_{\omega_n,\omega_n^\prime}\frac{\omega_n^2}{(\omega_n^2+\mathbf{k}^2)}\frac{\omega_n^{\prime2}}{(\omega_n^{\prime2}+\mathbf{k}^{\prime2})},\notag\\
    &=\sum_{\omega_n,\omega_n^\prime}\left[1-\frac{\mathbf{k}^2}{(\omega_n^2+\mathbf{k}^2)}\right]\left[1-\frac{\mathbf{k}^{\prime2}}{(\omega_n^{\prime2}+\mathbf{k}^{\prime2})}\right]\notag\\
    &=\sum_{\omega_n,\omega_n^\prime}\mathbf{1}-\sum_{\omega_n,\omega_n^\prime}\mathbf{1}\cdot \frac{\mathbf{k}^2}{\omega_n^2+\mathbf{k}^2}\notag\\
    &-\sum_{\omega_n,\omega_n^\prime}\mathbf{1}\cdot \frac{\mathbf{k}^{\prime2}}{\omega_n^{\prime 2}+\mathbf{k}^{\prime2}}+\sum_{\omega_n,\omega_n^\prime}\frac{\mathbf{k}^2}{\omega_n^2+\mathbf{k}^2}\frac{\mathbf{k}^{\prime2}}{\omega_n^{\prime 2}+\mathbf{k}^{\prime2}},
\end{align}
where the first three terms vanish after vacuum subtraction, for the same
reason as above.
The only nonzero contribution comes from the last part of Eq. \eqref{m1}, and the result is
\begin{align}
    \frac{\mathbf{k}^2}{2|\mathbf{k}|T}\left[1+2n_B(|\mathbf{k}|)\right]\frac{\mathbf{k}^{\prime2}}{2|\mathbf{k}|^\prime T}\left[1+2n_B(|\mathbf{k}^\prime|)\right],
\end{align}
using the standard Matsubara sum \eqref{MS1}, where
\(n_B(E_k)\equiv\frac{1}{e^{\beta E_k}-1}\) is the Bose distribution.
For the second part of the numerator, we use isotropic tensor reduction. First,
\begin{equation}
(\mathbf{k}\cdot\mathbf{k}')^2
=
(\delta^{ij}k_i k'_j)(\delta^{mn}k_m k'_n)
=
\delta^{ij}\delta^{mn}\,k_i k_m\,k'_j k'_n .
\end{equation}
Then
\begin{align}
&\int \frac{d^3\mathbf{k}\,d^3\mathbf{k'}}{(2\pi)^6}
\sum_{\omega_n,\omega_n'}
\frac{(\mathbf{k}\cdot\mathbf{k}')^2}{(\omega_n^2+\mathbf{k}^2)(\omega_n'^2+\mathbf{k'}^2)}
\nonumber\\
&=
\delta^{ij}\delta^{mn}\!\!\sum_{\omega_n,\omega_n'}\!\!\!
\left(
\int \frac{d^3\mathbf{k}}{(2\pi)^3}\,
\frac{k_i k_m}{\omega_n^2+\mathbf{k}^2}
\right)
\!\!
\left(
\int \frac{d^3\mathbf{k'}}{(2\pi)^3}\,
\frac{k'_j k'_n}{\omega_n'^2+\mathbf{k'}^2}
\right).
\end{align}
For any rotationally invariant function $H(\mathbf{k}^2)$, one has
\begin{equation}
\int d^3\mathbf{k}\, k_i k_j\, H(\mathbf{k}^2)
=
\frac{\delta_{ij}}{3}\int d^3\mathbf{k}\, \mathbf{k}^2\, H(\mathbf{k}^2),
\end{equation}
since the left-hand side is an isotropic rank-two tensor and must therefore be proportional to $\delta_{ij}$; contracting both sides with $\delta^{ij}$ fixes the coefficient to be $1/3$. Applying this identity to both momentum integrals, we obtain
\begin{align}
&\int \frac{d^3\mathbf{k}\,d^3\mathbf{k'}}{(2\pi)^6}
\sum_{\omega_n,\omega_n'}
\frac{(\mathbf{k}\cdot\mathbf{k}')^2}{(\omega_n^2+\mathbf{k}^2)(\omega_n'^2+\mathbf{k'}^2)}
\nonumber\\
&=
\frac{1}{9}\,
\delta^{ij}\delta^{mn}\delta_{im}\delta_{jn}
\int \frac{d^3\mathbf{k}\,d^3\mathbf{k'}}{(2\pi)^6}
\sum_{\omega_n,\omega_n'}
\frac{\mathbf{k}^2 \mathbf{k'}^2}{(\omega_n^2+\mathbf{k}^2)(\omega_n'^2+\mathbf{k'}^2)}
\nonumber\\
&=
\frac{1}{3}
\int \frac{d^3\mathbf{k}\,d^3\mathbf{k'}}{(2\pi)^6}
\sum_{\omega_n,\omega_n'}
\frac{\mathbf{k}^2 \mathbf{k'}^2}{(\omega_n^2+\mathbf{k}^2)(\omega_n'^2+\mathbf{k'}^2)} ,
\end{align}
where we used $\delta^{ij}\delta^{mn}\delta_{im}\delta_{jn}=3$. Using the same
Matsubara sum \eqref{MS1},
\begin{align}
&\sum_{\omega_n,\omega_n'}
\frac{\mathbf{k}^2}{\omega_n^2+\mathbf{k}^2}\,
\frac{\mathbf{k'}^2}{\omega_n'^2+\mathbf{k'}^2}\notag\\
&=
\frac{\mathbf{k}^2}{2|\mathbf{k}|T}\,[1+2n_B(|\mathbf{k}|)]\,
\frac{\mathbf{k'}^2}{2|\mathbf{k'}|T}\,[1+2n_B(|\mathbf{k'}|)] .
\end{align}
we obtain from Eq.~\eqref{f1-2}
\begin{align}
    f_{(1)}
    =&-\frac{16c}{M^4}T^2\int\frac{d^3\mathbf{k}\,d^3\mathbf{k^\prime}}{(2\pi)^6}\left(\frac{|\mathbf{k}||\mathbf{k}^\prime|}{4T^2}+\frac{1}{3}\frac{|\mathbf{k}||\mathbf{k}^{\prime}|}{4T^2}\right)\notag\\
    &\times\left[1+2n_B(|\mathbf{k}|)\right]\left[1+2n_B(|\mathbf{k}^\prime|)\right],
\end{align}
i.e.
\begin{align}
    f_{(1)}
    =&-\frac{16c}{3M^4}T^2\int\frac{d^3\mathbf{k}\,d^3\mathbf{k^\prime}}{(2\pi)^6}\frac{|\mathbf{k}||\mathbf{k}^\prime|}{T^2}\notag\\
    &\times\left[1+2\frac{1}{e^{\beta|\mathbf{k}|}-1}\right]\left[1+2\frac{1}{e^{\beta|\mathbf{k}^\prime|}-1}\right].
\end{align}
Then by the same argument of Eq. \eqref{f1-2}, only the purely thermal piece remains:
\begin{align}
    f_{(1)}
    &=-\frac{16c}{3M^4}\int\frac{d^3\mathbf{k}\,d^3\mathbf{k^\prime}}{(2\pi)^6}\frac{4|\mathbf{k}||\mathbf{k}^\prime|}{\left[e^{\beta|\mathbf{k}|}-1\right]\left[e^{\beta|\mathbf{k}^\prime|}-1\right]}\notag\\
    &=-\frac{16c}{3M^4\pi^4}\int dk\,dk^\prime\frac{k^3 k^{\prime3}}{\left[e^{\beta k}-1\right]\left[e^{\beta k^\prime}-1\right]}\notag\\
    &=-\frac{16c\pi^4}{675M^4\beta^8}.
\end{align}
Finally, the free energy density, including both the leading-order and first-order contributions, is given by
\begin{align}
    f=f_{(0)}+f_{(1)}=-\frac{\pi^2T^4}{90}-\frac{{16\,}c\pi^4T^8}{{675}M^4}.
\end{align}
At this point, having incorporated the leading and first-order terms, one might naturally ask whether higher-order corrections, such as the second-order term \( f_{(2)} \), should also be considered. To assess this, we estimate the magnitude of subsequent terms in the perturbative expansion. For example,
\begin{align}
f_{(2)} &\sim \frac{c^2}{M^8} T^{12} \sim c^2 \left( \frac{T}{M} \right)^8 T^4, \\
f_{(3)} &\sim \frac{c^3}{M^{12}} T^{16} \sim c^3 \left( \frac{T}{M} \right)^{12} T^4,
\end{align}
and so on. Each term is increasingly suppressed by higher powers of the expansion parameter \( (T/M)^4 \), whether arising from higher-loop corrections or higher-dimensional operators.


The entropy density is therefore
\begin{align}
    s=-\frac{\partial f}{\partial T}=\frac{2\pi^2T^3}{45}+\frac{128\,c\pi^4T^7}{675M^4}.
\end{align}
For our positivity analysis, we work with thermal entropy density (entropy per
unit volume) in the canonical ensemble at fixed $T$ and $V$, and compare two
theories: the low-energy EFT obtained after integrating out the heavy sector,
which includes the induced higher-derivative interaction proportional to $c$,
and the free light theory obtained by removing the heavy sector and setting
$c=0$.

The physical requirement we impose is that, in a consistent ultraviolet
completion, integrating out additional microscopic degrees of freedom should not
reduce the thermal entropy of the resulting low-energy description in the regime
$m \ll T \ll M$. Importantly, this statement is not a claim of monotonicity under
arbitrary interactions at fixed $T$, but rather a comparison between two
different theories: one obtained from a UV-complete model by integrating out a
heavy field, and one in which that heavy sector is absent altogether.

Evaluating the entropy density in the effective theory perturbatively in
$T/M \ll 1$, we find
\begin{equation}
    s_{\mathrm{EFT}}(T)
    = s_{\mathrm{free}}(T)
    + \frac{128\,\pi^4}{675}\,\frac{c\,T^7}{M^4}
    + \mathcal{O}\!\left(\frac{T^{11}}{M^8}\right) .
\end{equation}
Requiring that the entropy density of the EFT exceed that of the free light
theory,
\begin{equation}
    s_{\mathrm{EFT}}(T) > s_{\mathrm{free}}(T)
    \qquad (m \ll T \ll M),
\end{equation}
immediately implies
\begin{equation}
    c > 0 .
\end{equation}

This criterion can be viewed as a thermodynamic consistency condition on the
low-energy description induced by heavy physics, rather than a general statement
about interacting systems at fixed temperature.
\section{Illustrative ultraviolet completions}
\subsection{A global $U(1)$ UV completion }
\label{app:globalU1completion}

To make the comparison leading to $c>0$ more explicit, it is useful to exhibit a
simple UV completion in which the Goldstone EFT arises by integrating out a
heavy field. A minimal example is the global $U(1)$ linear sigma model,
\begin{equation}
\mathcal{L}_{\rm UV}
= -|\partial_\mu \Phi|^2
-\lambda\left(|\Phi|^2-\frac{v^2}{2}\right)^2,
\qquad \lambda>0,
\label{eq:UV_sigma}
\end{equation}
where $\Phi$ is a complex scalar and the vacuum spontaneously breaks the global
$U(1)$ symmetry. Parametrizing fluctuations in polar variables,
\begin{equation}
\Phi(x)=\frac{v+h(x)}{\sqrt{2}}\,e^{i\phi(x)/v},
\label{eq:polar}
\end{equation}
the Lagrangian becomes
\begin{align}
\mathcal{L}_{\rm UV}
= -\frac12(\partial h)^2
-\frac12&\left(1+\frac{h}{v}\right)^2(\partial\phi)^2
-\frac12 m_h^2 h^2
+\cdots,\\
m_h^2&=2\lambda v^2,
\label{eq:UV_expand}
\end{align}
where $h$ is the heavy radial mode and $\phi$ is the (massless) Goldstone boson.
At energies and temperatures well below the radial mass, $E,T\ll m_h$, we may
integrate out $h$ to obtain a local EFT for $\phi$. At tree level, solving the
equation of motion for $h$ gives
\begin{equation}
h \simeq -\frac{1}{v\,m_h^2}(\partial\phi)^2
+\mathcal{O}\!\left(\frac{\partial^4}{m_h^4}\right),
\label{eq:hEOM}
\end{equation}
which upon substitution into \eqref{eq:UV_expand} yields
\begin{equation}
\mathcal{L}_{\rm EFT}
= -\frac12(\partial\phi)^2
+ \frac{1}{2v^2 m_h^2}\,(\partial\phi)^4
+ \mathcal{O}\!\left(\frac{\partial^6}{m_h^4}\right).
\label{eq:EFT_match}
\end{equation}
Matching to the form used in the main text,
\begin{equation}
\mathcal{L}_{\rm EFT}
=-\frac12(\partial\phi)^2+\frac{c}{M^4}(\partial\phi)^4+\cdots,
\end{equation}
and identifying $M\equiv m_h$, we obtain
\begin{equation}
\frac{c}{M^4}=\frac{1}{2v^2 m_h^2}
\quad\Rightarrow\quad
c=\frac{m_h^2}{2v^2}=\lambda>0.
\label{eq:cpos_globalU1}
\end{equation}
Thus, in this explicit UV completion the positivity of $c$ follows directly from
the stability condition $\lambda>0$ of the scalar potential.

\textbf{Thermal free energy and entropy at $T\ll M$.}
In the full UV theory \eqref{eq:UV_sigma}, the heavy radial mode contributes to
thermal quantities only through Boltzmann-suppressed terms $\sim e^{-m_h/T}$.
Therefore, for $T\ll m_h$ the temperature-dependent part of the free energy is
dominated by the light Goldstone sector and is captured by the EFT
\eqref{eq:EFT_match}. Expanding perturbatively in $T/m_h\ll1$ reproduces the same
leading correction found in the main text,
\begin{equation}
f_{\rm EFT}(T)
= f_{\rm free}(T)
-\frac{16\pi^4}{675}\,\frac{c\,T^8}{M^4}
+ \mathcal{O}\!\left(\frac{T^{12}}{M^8}\right),
\end{equation}
and hence
\begin{equation}
s_{\rm EFT}(T)-s_{\rm free}(T)
= +\frac{128\pi^4}{675}\,\frac{c\,T^7}{M^4}
+ \mathcal{O}\!\left(\frac{T^{11}}{M^8}\right).
\label{eq:entropy_shift_globalU1}
\end{equation}
Requiring that the low-energy description obtained by integrating out the heavy
field have larger thermal entropy density than the free light theory in the
regime $m\ll T\ll M$ then implies $c>0$, in agreement with the UV matching
\eqref{eq:cpos_globalU1}. We emphasize that this is a comparison between two
different theories (with and without the heavy sector), rather than a general
monotonicity statement for arbitrary interactions at fixed temperature.

\subsection{$\lambda\phi^4$ interaction and its UV completion}
\label{subsec:phi4_UV}

We begin by considering a scalar field theory in four dimensions with a quartic
self-interaction,
\begin{equation}
\mathcal{L}
= -\frac12(\partial_\mu\phi)^2 - \lambda\,\phi^4 ,
\label{eq:phi4_Lagrangian}
\end{equation}
where we follow the sign convention appropriate to our thermodynamic analysis.
With this convention, stability of the vacuum requires
\begin{equation}
\lambda > 0 .
\end{equation}
This sign choice does not contradict our entropy bound, since the operator
$\phi^4$ is marginal in four dimensions and should be interpreted as a
self-interaction of the light field, rather than as an operator generated by
integrating out heavy degrees of freedom. In general, such marginal interactions
are not constrained by our entropy-based argument.

However, in certain restricted situations the quartic interaction can arise
from integrating out a heavy field. In such cases, the sign of $\lambda$ is not
arbitrary but is fixed by the structure of the ultraviolet completion. A simple
toy model illustrating this point is provided by tree-level exchange of a heavy
scalar field.

Consider a UV theory in which the light scalar $\phi$ interacts with a heavy
scalar $\Phi$ of mass $M$ through a cubic coupling,
\begin{equation}
\mathcal{L}_{\rm UV}
= -\frac12(\partial_\mu\phi)^2
  -\frac12(\partial_\mu\Phi)^2
  -\frac12 M^2 \Phi^2
  -\frac{g}{2}\,\phi^2 \Phi .
\label{eq:phi4_UV}
\end{equation}
At energies well below the heavy mass scale, $E\ll M$, the heavy field $\Phi$ can
be integrated out. At tree level, this amounts to solving its classical equation
of motion,
\begin{equation}
(\Box + M^2)\Phi = -\frac{g}{2}\,\phi^2 ,
\end{equation}
which in the low-energy limit reduces to
\begin{equation}
\Phi(x) \simeq -\frac{g}{2M^2}\,\phi^2(x) .
\end{equation}
Substituting this solution back into the Lagrangian \eqref{eq:phi4_UV}, one
obtains the effective quartic interaction for $\phi$,
\begin{equation}
V_{\rm eff}(\phi)
\simeq
\frac12 M^2\!\left(-\frac{g}{2M^2}\phi^2\right)^2
+ \frac{g}{2}\phi^2\!\left(-\frac{g}{2M^2}\phi^2\right)
= -\frac{g^2}{8M^2}\,\phi^4 .
\label{eq:phi4_matching}
\end{equation}
Comparing with \eqref{eq:phi4_Lagrangian}, we identify the Wilson coefficient
\begin{equation}
\lambda = -\frac{g^2}{8M^2} < 0 .
\end{equation}
Since $g^2>0$ and $M^2>0$, the quartic coupling generated by tree-level exchange
of a heavy scalar is necessarily negative. This reflects the general fact that
single-boson exchange mediates an attractive interaction, corresponding to a
negative contribution to the effective potential.

In this sense, when a $\lambda\phi^4$ interaction arises from integrating out
heavy degrees of freedom, its sign is fixed and agrees with the expectation from
the underlying UV physics. Our entropy bound is consistent with this structure:
it constrains leading higher-dimensional operators generated by heavy fields, while
allowing marginal self-interactions such as $\phi^4$ to have either sign unless
they are tied to a specific UV completion.

\section{Connection to unitarity and causality}
\label{sec:causality_unitarity}

Most existing positivity bounds are derived from the analytic structure of
$2\to2$ scattering amplitudes, where fundamental principles such as unitarity
and causality play a central role. Unitarity implies the optical theorem, which
enforces the non-negativity of the imaginary part of forward scattering
amplitudes, while causality leads to analyticity properties of amplitudes in the
complex energy plane. Combined with dispersion relations, these properties
translate into positivity constraints on Wilson coefficients in the low-energy
effective action.

The entropy-based bound derived in this work follows a different logical route
and does not rely directly on scattering amplitudes. Instead, it is based on
thermodynamic considerations and general properties of quantum entropy. It is
therefore appropriate to clarify the limited and indirect sense in which
unitarity and causality enter the discussion.

\paragraph{Unitarity and entropy inequalities.}
The entropy inequalities employed in this work, in particular the
Araki--Lieb inequality~\cite{Araki:1970ba,Witten:2018zva}, are kinematic
statements about von Neumann entropies of density matrices. They hold for any
bipartite quantum system and do not require unitary time evolution or specific
dynamical assumptions. In particular, unitarity does not play a role in the
derivation of the Araki--Lieb inequality itself.

In our context, unitarity is assumed only at a minimal level: namely, that the
quantum theory admits a well-defined density matrix with positive spectrum and
conserved trace,
\begin{equation}
\frac{d}{dt}\,\mathrm{Tr}(\rho)=0 .
\end{equation}
These basic consistency conditions ensure that von Neumann entropy is
well-defined and that general entropy inequalities can be meaningfully applied.
Beyond this assumption of probability conservation and positivity, no further
use of dynamical unitarity is made in our argument.

\paragraph{Causality and relativistic consistency.}
Causality enters our analysis only insofar as the thermal quantum field theory
under consideration is assumed to be a consistent relativistic QFT. In the
Euclidean formulation used to compute the thermal partition function, there is
no notion of real-time evolution or signal propagation. The connection to
causality arises only upon analytic continuation to Lorentzian signature.

In axiomatic quantum field theory, relativistic causality is encoded in the
principle of locality, formulated for example in the Haag--Kastler framework
as the commutativity of observables at spacelike separation~\cite{Haag:1963dh},
\begin{equation}
[\phi(x),\phi(y)]=0
\qquad \text{for} \qquad (x-y)^2>0 .
\end{equation}
This microcausality condition follows from the analytic structure of correlation
functions and is consistent with the reconstruction of Lorentzian QFT from
Euclidean correlation functions satisfying the Osterwalder--Schrader axioms,
including reflection positivity~\cite{Osterwalder:1973dx,Osterwalder:1974tc}.

In practical terms, this structure is reflected in the standard
\(i\epsilon\) prescription of the Feynman propagator,
\begin{equation}
G_F(x-y)=\int\frac{d^4p}{(2\pi)^4}\,
\frac{i\,e^{-ip\cdot(x-y)}}{p^2-m^2+i\epsilon},
\end{equation}
which specifies how Lorentzian correlators are obtained as boundary values of
Euclidean correlation functions. In the present work, causality plays no direct
role in deriving the entropy inequality, but it underlies the assumption that
the Euclidean thermal path integral admits a consistent continuation to a
relativistic quantum theory.

\paragraph{Summary.}
In summary, the entropy-based positivity bound presented here does not rely on
unitarity and causality as direct inputs in the way amplitude-based approaches
do. Instead, these principles enter only indirectly, through the assumption that
the underlying quantum field theory is internally consistent, admits a
well-defined thermal partition function, and satisfies general quantum
information inequalities. From this perspective, our result should be viewed as
complementary to existing positivity bounds derived from scattering amplitudes,
rather than as an alternative derivation of them.
\section{Conclusion and discussion}\label{ConclusionSection}
In this work we computed the thermal entropy of a shift-symmetric scalar EFT and
showed that thermodynamic consistency yields a nontrivial sign constraint on the
effective interactions. In particular, in the regime $m\ll T\ll M$, the entropy
comparison between the heavy-induced EFT and the free light theory implies that
the leading dimension-8 Wilson coefficient is strictly positive.

The derivation is based on finite-temperature field theory and entropy
inequalities, rather than on scattering amplitudes and dispersion relations.
Viewed this way, the result provides a complementary perspective on EFT
consistency and clarifies how thermodynamic data can encode constraints usually
associated with analyticity, unitarity, and causality.

The same strategy can be extended in several directions, including EFTs with
fermions, gauge fields, and gravity. These settings are often difficult for
standard amplitude-based positivity bounds because of massless exchanges and, in
some backgrounds, the lack of a conventional S-matrix. Since our method does not
depend on a flat-space S-matrix, it may provide a useful alternative diagnostic
in such cases.


After posting v1 of this manuscript, we became aware of a closely related preprint by Fernández-Sarmiento, Penco, and Rosen~\cite{Fernandez-Sarmiento:2025tpy} through correspondence from Riccardo Penco. Their analysis formulates an entropy-positivity condition in terms of a negative thermal shift of the grand potential and tests it across several systems, together with a discussion of regimes where the bound may not apply. Our approach starts from a different argument: by beginning with the Araki-Lieb inequality and an explicit leading-order thermal QFT computation, we demonstrate this condition in a Lorentz-invariant scalar EFT and obtain the positive sign of the leading dimension-8 operator. We also outline the assumptions of our derivation and its regime of validity within this setup.

\section*{Acknowledgements}
We are grateful to Anna Tokareva for a careful reading of the manuscript and for insightful comments. Y.X. also thanks Longqi Shao and Shuang-Yong Zhou for inspiring discussions, and colleagues at the Amplitudes 2025 conference (Seoul National University, June 16–20, 2025) for helpful feedback on a preliminary version of this work presented as a poster. Finally, we thank Riccardo Penco, Lucas Fernandez Sarmiento, and Brando Bellazzini for useful comments and discussions.

\appendix
\section{Useful Matsubara sum formulas}
In this appendix we collect and derive several Matsubara sum identities used in the main text. We focus on bosonic Matsubara frequencies
\begin{align}
    \omega_n=2\pi nT,\quad T=\beta^{-1}, \quad n\in \mathbb{Z},
\end{align}
which arise naturally in the imaginary-time formalism of finite-temperature field theory.

Throughout this appendix we consider a massless scalar field for simplicity, since this is the case relevant for the EFT discussed in the main text.

\subsection{The contour integral method}
The standard technique for evaluating sums of the form \(S=T\sum^\infty_{n=-\infty}g(i\omega_n)\) is to convert the sum into a contour integral in the complex energy plane ($z$-plane). We introduce the Bose-Einstein distribution function as a weighting function:
\begin{align}
    n_B(z)=\frac{1}{e^{\beta z}-1}.
\end{align}
The function \(n_B(z)\) has simple poles strictly on the imaginary axis at \(z=i\omega_n=i\frac{2\pi n}{\beta}\) with residue T:
\begin{align}
    {\rm Res}\left[n_{B}(z),\,z=i\omega_n\right]=\lim_{z\to i \omega_n}\frac{z-i\omega_n}{e^{\beta z}-1}=\frac{1}{\beta e^{i 2 \pi n}}=T.
\end{align}
By Cauchy's residue theorem, the sum can be written as a contour integral enclosing the imaginary axis:
\begin{align}
    T \sum_{n=-\infty}^{\infty} g\left(i \omega_n\right)=\oint_{\mathcal{C}} \frac{d z}{2 \pi i} g(z) n_B(z).
\end{align}
where the contour \(\mathcal{C}\) consists of small counter-clockwise circles around each pole \(i\omega_n\). Assuming \(g(z)\) is analytic except for specific poles on the real axis (and decays sufficiently fast at infinity), we can deform the contour \(\mathcal{C}\) into two integration lines running parallel to the imaginary axis (one slightly to the right, one slightly to the left) and effectively wrap them around the physical poles of \(g(z)\).

This leads to the ``Master Formula'' for Matsubara summation:
\begin{equation}
    T \sum_{n=-\infty}^{\infty} g(i\omega_n) = - \sum_{z_k} \text{Res} \left[ g(z) n_B(z) \right]_{z=z_k}
\end{equation}
where $z_k$ are the poles of the function $g(z)$.

\subsection{Summation for the propagator}
We compute the sum appearing in the perturbation diagrams:
\begin{align}
    S_1=T \sum_{n=-\infty}^{\infty} \frac{1}{\omega_n^2+\mathbf{k}^2},
\end{align}
Here, \(g\left(i \omega_n\right)=\frac{1}{-\left(i \omega_n\right)^2+\mathbf{k}^2}=\frac{1}{-z^2+\mathbf{k}^2}=-\frac{1}{z^2-\mathbf{k}^2}\). The function \(g(z)=-\frac{1}{(z-|\mathbf{k}|)(z+|\mathbf{k}|)}\) has two simple poles on the real axis at \(z=\pm |\mathbf{k}|\). Applying the residue formula:
\begin{align}
S_1 & =-\left(\operatorname{Res}\left[\frac{-n_B(z)}{z^2-\mathbf{k}^2}\right]_{z=|\mathbf{k}|}+\operatorname{Res}\left[\frac{-n_B(z)}{z^2-\mathbf{k}^2}\right]_{z=-|\mathbf{k}|}\right) \\
& =\frac{n_B(|\mathbf{k}|)}{2 |\mathbf{k}|}+\frac{n_B(-|\mathbf{k}|)}{-2 |\mathbf{k}|} \\
& =\frac{1}{2 |\mathbf{k}|}\left[n_B(|\mathbf{k}|)-n_B(-|\mathbf{k}|)\right]
\end{align}
Using the identity \(n_B(-|\mathbf{k}|)=-(1+n_B(|\mathbf{k}|))\), we obtain the standard result:
\begin{align}
    T \sum_{n=-\infty}^{\infty} \frac{1}{\omega_n^2+\mathbf{k}^2}=\frac{1}{2 |\mathbf{k}|}\left[1+2 n_B(|\mathbf{k}|)\right].\label{MS1}
\end{align}
The two terms correspond to different physics: The first term \(1/(2k)\) corresponds to the zero-temperature (vacuum) contribution. The second term \(n_B(k)/k\) represents the thermal excitation of bosonic modes. In thermal calculations, one often subtracts the vacuum piece, keeping only the \(n_B(k)\) terms' contribution.

\subsection{Sum of logarithms: \(\sum_n \ln\left[\left(\omega_n^2+\mathbf{k}^2\right)\beta^2\right]\)}
In the evaluation of the free partition function, one encounters sums of the form:
\begin{align}
    S_2=\sum_{n=-\infty}^{\infty} \ln \left[\left(\omega_n^2+\mathbf{k}^2\right)\beta^2\right],
\end{align}
Since \(\ln(z)\) has a branch cut rather than simple poles, the direct residue method is subtle. Instead, differentiating with respect to \(|\mathbf{k}|\) gives
\begin{align}
    \frac{d S_2}{d |\mathbf{k}|}=\sum_n \frac{2 |\mathbf{k}|}{\omega_n^2+\mathbf{k}^2}=2 |\mathbf{k}| \sum_n \frac{1}{\omega_n^2+\mathbf{k}^2}.
\end{align}
Using Eq.~\eqref{MS1},
\begin{align}
    \frac{d S_2}{d |\mathbf{k}|}=2 |\mathbf{k}| \cdot \beta \frac{1}{2 |\mathbf{k}|}\left[1+2 n_B(|\mathbf{k}|)\right]=\beta\left[1+2 n_B(|\mathbf{k}|)\right] .
\end{align}
Integrating with respect to \(|\mathbf{k}|\)
\begin{align}
    S_2(|\mathbf{k}|)=\beta |\mathbf{k}|+2 \int d |\mathbf{k}^{\prime}| \beta n_B\left(|\mathbf{k}^{\prime}|\right)+\text { const. }
\end{align}
Using
\begin{align}
    \beta n_B(|\mathbf{k}|)=\frac{d}{d |\mathbf{k}|} \ln \left(1-e^{-\beta |\mathbf{k}|}\right),
\end{align}
we obtain
\begin{align}
    \sum_n \ln \left[\left(\omega_n^2+\mathbf{k}^2\right)\beta^2\right]=\beta |\mathbf{k}|+2 \ln \left(1-e^{-\beta |\mathbf{k}|}\right)+\text { const } .\label{MS2}
\end{align}
The additive constant is independent of \(|\mathbf{k}|\) and contributes only to a temperature-independent vacuum term, which can be dropped when computing thermodynamic quantities.

\section{Poisson summation and thermal sum decomposition}\label{Poisson}
In this appendix we derive the Poisson summation formula and show how it leads to the decomposition of Matsubara sums into vacuum and thermal contributions. This is used in the main text to justify that constant sum-integrals contain only vacuum contributions.


\subsection{Derivation of the Poisson summation formula}
Let \(f(x)\) be a function on the real line with Fourier transform
\begin{align}
    \tilde{f}(k)=\int_{-\infty}^{\infty} d x e^{-i k x} f(x).
\end{align}
We define a sum:
\begin{align}\label{defOfP}
    P(x)\equiv \sum_{n\in\mathbb{Z}}f(x+na).
\end{align}
It is immediate that \(P(x)\) is periodic with period \(a\):
\(P(x+a)=P(x)\). Any sufficiently regular periodic function admits a Fourier
series expansion, so
\begin{align}\label{seriesOfP}
    P(x)=\sum_{m \in \mathbb{Z}} c_m e^{i \frac{2 \pi m}{a} x},
\end{align}
with Fourier coefficients
\begin{align}\label{defOfcm}
    c_m=\frac{1}{a} \int_0^a d x P(x) e^{-i \frac{2 \pi m}{a} x}.
\end{align}
Substituting Eq. \eqref{defOfP} into Eq. \eqref{defOfcm} gives
\begin{align}
    c_m=\frac{1}{a} \int_0^a d x \sum_{n \in \mathbb{Z}} f(x+n a) e^{-i \frac{2 \pi m}{a} x} .
\end{align}
Equivalently,
\begin{align}
    c_m&=\frac{1}{a} \int_0^a d x \sum_{n \in \mathbb{Z}} f(x+n a) e^{-i \frac{2 \pi m}{a} x}\notag\\
    &=\frac{1}{a} \sum_{n \in \mathbb{Z}} \int_0^a d x f(x+n a) e^{-i \frac{2 \pi m}{a} x}\notag\\
    &=\frac{1}{a} \int_{-\infty}^{\infty} d x f(x) e^{-i \frac{2 \pi m}{a} x}\notag\\
    &=\frac{1}{a}\tilde{f}\left(\frac{2\pi m}{a}\right).
\end{align}
Substituting into Eq. \eqref{seriesOfP} gives
\begin{align}
    P(x) = \frac{1}{a} \sum_{m \in \mathbb{Z}} \tilde{f}\!\left(\frac{2\pi m}{a}\right) e^{i \frac{2\pi m}{a} x}.
\end{align}
Evaluating at \(x=0\) yields the Poisson summation formula,
\begin{align}\label{PoissonSumFormula}
    \sum_{n \in \mathbb{Z}} f(na) = \frac{1}{a} \sum_{m \in \mathbb{Z}} \tilde{f}\!\left(\frac{2\pi m}{a}\right).
\end{align}


\subsection{Constant integrand and absence of thermal correction}
To analyze the constant integrand, we introduce a regulator
\begin{align}
    F_\alpha(\omega)=e^{-\alpha|\omega|},\quad\alpha>0,
\end{align}
and take \(\alpha\to 0^+\) at the end. Using Eq. \eqref{PoissonSumFormula}
\begin{align}
    T \sum_{n \in \mathbb{Z}} F_{\alpha}(\omega_n)
= \sum_{m \in \mathbb{Z}} \int_{-\infty}^{\infty} \frac{d\omega}{2\pi}
\, e^{i m \beta \omega} e^{-\alpha |\omega|}.
\end{align}
For \(m=0\),
\begin{align}
    \int \frac{d\omega}{2\pi}e^{-\alpha|\omega|},
\end{align}
the integrand does not depend on \(\beta\) and therefore produces only a temperature-independent contribution. In thermal field theory, such contributions correspond to vacuum normalization and are removed by the standard subtraction of the zero-temperature free energy.

While the \(m\neq 0\) terms encode genuine thermal corrections, and the integral evaluates to
\begin{align}
    \frac{1}{\pi}\frac{\alpha}{\alpha^2+(m\beta)^2},
\end{align}
which vanishes as \(\alpha\to 0^+\). Therefore the thermal correction vanishes:
\begin{align}
    \Delta_T[1]=0, \quad \Rightarrow \quad\left[T \sum_n-\int \frac{d \omega}{2 \pi}\right] 1=0 .
\end{align}
Thus constant sum-integrals contain only vacuum contributions and are removed by vacuum subtraction in thermal free energy.

\bibliography{Refs-Entropy}

\end{document}